\documentstyle[amssymb,prl,aps,twocolumn,epsf,floats]{revtex}
\newcommand{\be}{\begin{equation}}
\newcommand{\ee}{\end{equation}}
\newcommand{\bea}{\begin{eqnarray}}
\newcommand{\eea}{\end{eqnarray}}

\begin{document}

\twocolumn[\hsize\textwidth\columnwidth\hsize\csname @twocolumnfalse\endcsname 

\title {Spectral function and conductivity in the normal state of the 
cuprates: a spin fluctuation study.}
\author{R. Haslinger, Andrey V. Chubukov, 
and Ar. Abanov} 
\address{ 
$^1$ Department of Physics, University of Wisconsin, Madison, WI 53706} 
\date{\today} 
\draft 
\maketitle  
\begin{abstract} 
We study the spectral function $A_k (\omega)$ and the optical conductivity 
$\sigma_1 (\omega)$ for a system of fermions interacting with their collective spin 
fluctuations (the spin-fermion model), and apply the results to optimally doped 
cuprates in the normal state. We show
 that there is no qualitative distinction between hot and cold regions in the 
 Brillouin zone - in both cases, there exists a wide range of frequencies where 
 the width of the peak in $A_k (\omega)$ scales linearly with $\omega$. 
%The slope, however, depends on $k$ and is the
%largest near hot spots. 
We demonstrate that $\sigma_1 (\omega)$ is inversely linear in $\omega$ up to very
 high frequencies. We argue that these results 
agree {\it quantitatively} with the
photoemission and optical data for $Bi2212$.
\end{abstract} 
\pacs{PACS numbers:71.10.Ca,74.20.Fg,74.25.-q} 

] \narrowtext 
The observed  discrepancies between the
 normal state properties of the cuprates and the 
predictions  of Fermi liquid theory  continue to attract a lot of
attention from the condensed-matter community. 
 According to a Fermi liquid theory,
the behavior of all fermionic systems at sufficiently low energies is universal 
and is governed by the fact that the quasiparticle 
damping near the Fermi surface scales as $\omega^2$ or $T^2$ (whichever is larger).
  This should  give rise to the 
$\omega^2$ ($T^2$) behavior of the width of the photoemission peak,  
to $1/\omega^2$ behavior of the 
optical conductivity $\sigma_1 (\omega)$ at $T=0$,  and to the $T^2$ behavior 
of the resistivity $\rho (T)$.

Traces of Fermi liquid behavior have been observed in strongly overdoped cuprates. 
For smaller dopings, however, the deviations from the Fermi liquid behavior 
become  substantial, and even at
optimal doping, the system behavior in all experimentally accessable frequency ranges
is qualitatively different from that in a Fermi liquid~\cite{and}. 
Amazingly enough, this non-Fermi liquid behavior can,  to good accuracy, 
be described by simple linear functions of frequency and temperature. In particular, 
the width of the ARPES peak  scales with $\omega$ for a wide range of frequencies, 
%at least for momenta close to the zone diagonal,
the conductivity is inversely proportional to $\omega$, and the resistivity is linear in $T$.
To which extent this behavior survives in underdoped cuprates is unclear as below optimal doping, 
the normal state behavior is masked by the development of the pseudogap.
 
There are currently 
three qualitatively different phenomenological
 scenarios for the observed linearities at optimal doping. 
The first is a  marginal Fermi liquid scenario~\cite{mfl}. 
It assumes that near optimal doping, there exists a
quantum critical point of unknown origin, 
at which $\Sigma^{\prime \prime}_k (\omega)$ 
is linear in $\omega$ and independent of $k$.
The linearity of the resistivity and inverse optical
conductivity then follow from a conventional Drude theory.

The second is a  cold spot scenario~\cite{im}. 
It assumes that the quasiparticle lifetime is very anisotropic along the 
Fermi surface and preserves a Fermi liquid form only in a
narrow range near the zone diagonal where the 
 quasiparticle lifetime is the longest. The width of this range is assumed to
scale with $\omega$ (or $T$). Elementary manipulations show that the Fermi liquid region then 
yields the linear behavior  of resistivity and inverse conductivity.    

The third is a magnetic 
hot spot scenario which 
relates the 
linear behavior of the resistivity with the non-Fermi liquid behavior
due to strong spin-fluctuation scattering.
This scenario is questionable on general grounds as 
 the computation of conductivity 
implies averaging of the quasiparticle lifetime over the Fermi surface
~\cite{hr}, 
%(and hence is likely confined to the regions where the scattering is the 
%weakest),
 but still may be applicable by numerical reasons.  

In the present communication, we adopt a non-phenomenological approach and 
compute the ARPES lineshape, optical conductivity and resistivity for 
the spin-fermion model which describes  fermions interacting with their 
own collective spin degrees of freedom. 
Two of us argued in a series of recent publications~\cite{ac} that the 
strong coupling limit of this model 
captures the physics of the cuprates. 
We will use a previously obtained expression  for the fermionic self energy 
$\Sigma$ as an input, and 
compute the photoemission intensity for various $k$, 
optical conductivity and resistivity. 
 
The results we obtain  partly agree and partly disagree with each of  
the three phenomenological scenarious above. 
In agreement with the marginal Fermi liquid picture, we found that 
$\Sigma^{\prime \prime} (\omega)$ is linear in $\omega$ and in $T$ for  
a wide range of frequencies everywhere on the Fermi surface. 
This linearity in $\Sigma^{\prime \prime}$ causes the linear behavior 
of both the optical conductivity and 
the resistivity. In contradiction with the marginal Fermi liquid scenario
the linear behavior of $\Sigma^{\prime \prime} (\omega)$ in 
the spin-fermion model is not associated with a closeness to an
unknown  phase transition at optimal doping. Rather, it 
emerges as an intermediate 
asymptotics in the crossover regime between the two physically 
motivated limits: a Fermi liquid 
at the lowest frequencies, where 
$\Sigma^{\prime \prime} \propto \omega^2$
and a high frequency limit where the system is in the 
magnetic quantum-critical  regime, and
$\Sigma^{\prime \prime} \propto \sqrt{\omega}$~\cite{ac}. 
%This quantum-critical behavior is associated with the closeness to the 
%$T=0$ transition to the magnetically ordered state. 
The crossover between the two limits is governed by a single parameter - a 
momentum dependent typical spin relaxation frequency $\omega_{sf} (k)$. 
We find, however, that 
the crossover region is strikingly wide and covers the whole 
frequency range probed in the experiments.

In agreement with the cold spot scenario, we found that the amplitude 
of the scattering rate is anisotropic over the Fermi surface, and
is smallest near the zone diagonals. Acoordingly, 
(i) the slope of $\Sigma (\omega)$ is the smallest
and (ii) the crossover 
frequency $\omega_{sf} (k)$ is the largest for $k = k_{diag}$.
However, we found that
%Still, however, 
%for parameters which, we believe, are relevant 
%for optimally doped 
%cuprates, 
$\omega_{sf} (k_{diag})$ is very low, 
($ \leq 50 meV$), such that for 
$\omega$ and/or $\pi T$ probed in ARPES
and conductivity experiments, the system is outside the Fermi liquid regime
even for $k = k_{diag}$.
% 
%the quasiparticle damping is linear in $\omega$. Alternatively speaking, we find that for frequencies probed in the
%experiments, the quasiparticle damping is linear in $\omega$ everywhere on the Fermi surface, but the overall amplitude of the
%damping rate is the smallest near diagonals. 

Finally, in agreement with the hot spot theory, 
we find that the damping rate is indeed the strongest near the hot spots, and 
that there is a substantial, although not dominant, contribution 
to the conductivity from the hot regions.

We now turn to the calculations. The 
fermionic self-energy in the spin-fermion model has been obtained
before~\cite{ac} and is the starting point for our studies in this paper. 
We have 
\begin{equation}
\Sigma_k (\omega _{m})= i \pi T \lambda
~\sum_{n} \frac{\mbox{sign} \omega_n}{\sqrt{1 + \frac{|\omega_m -
 \omega_n|}{\omega_{sf}} + ({\tilde k} \xi)^2}}.
\label{z}
\end{equation}
where $\omega_m$ and $\omega_n$ are fermionic Matsubara frequencies.
Here $\lambda \sim \xi$
%$ = 3 {g^2 \chi_0}/(4 \pi v_f \xi^{-1})$ 
is the dimensionless effective coupling,
${\tilde k} = |{\bf k} - {\bf k}_{hs}|$ is the momentum 
deviation from a hot spot along the Fermi surface, and 
$\omega_{sf} = (3 \sin \phi_0/(16\pi)) v_F \xi^{-1}/\lambda$, where
$v_F$ is the Fermi velocity,
$\phi_0$ is the angle between the Fermi  velocities at 
$k$ and $k +Q$, and $\xi$ -- is the magnetic correlation length.

The physical meaning of $\omega_{sf}$ can be understood by analyzing Eq. (\ref{z}) at 
 $T=0$ and $k = k_{hs}$. In this limit, the frequency summation and 
 transformation to a real axis 
can be performed exactly and yields
 $ \Sigma_{k_{hs}} (\omega)=  2~\lambda \omega/(1+\sqrt{1-i |\omega|/\omega _{sf}})$.
A simple manipulation then shows that $\omega_{sf}$ is a crossover scale 
between a Fermi liquid behavior at 
$\omega \ll \omega_{sf}$ where 
 $\Sigma_{k_{hs}} (\omega) \approx   \lambda (\omega + i \omega |\omega|/(4\omega_{sf}))$, and the
quantum-critical $(\xi = \infty)$ behavior at $\omega \gg \omega_{sf}$ where 
$\Sigma_{k_{hs}} (\omega) \approx i  \mbox{sign} \omega (|\omega|~{\bar \omega})^{1/2}$,
and ${\bar \omega} = 4 \lambda^2 \omega_{sf}$
% = 9{g^2 \chi_0}/(8\pi)$ 
is independent on $\xi$. 

In  Fig. \ref{fig1}a,b we present plots of   
$\Sigma_{k_{hs}} (\omega, T)$. We computed
 $\Sigma^{\prime \prime} (\omega, T)$ 
by using the spectral representation for (\ref{z}), 
and then used a Kramers-Kronig transform to obtain 
$\Sigma^{\prime} (\omega, T)$. 
We see that at a given $T$ , $\Sigma^{\prime \prime} (\omega)$ 
is linear in $\omega$ at intermediate frequencies. 
At $T=0$, the 
linear regime extends  between $0.5 \omega_{sf}$ and $8 \omega_{sf}$. 
This linear behavior has been observed in numerical
studies~\cite{dev_kampf}. 
In Fig.\ref{fig1}c we see that $\Sigma^{\prime \prime}_{k_{hs}} (T)$  at fixed, small  $\omega$ is 
linear in $T$ above $100K$.
Theoretically, the linearity in $T$ can be understood as coming 
primarily from the scattering by thermal, classical spin fluctuations
(i.e., from the $n=m$ term in (\ref{z})).
% which obviously gives a linear in $T$ contribution to $\Sigma^{\prime \prime}$.
In Fig.\ref{fig1}d-f we present the results for the 
photoemission intensity $I_k (\omega) = A_k (\omega) n_F (\omega)$, where 
$A_k (\omega) = (1/\pi) \Sigma^{\prime \prime}_k (\omega)/((\omega - \epsilon_k 
+ \Sigma^{\prime}_k (\omega))^2 + ( \Sigma^{\prime \prime} (\omega))^2)$ is
the quasiparticle spectral function.
Fig. 1d shows  $I_k (\omega)$ for various $\epsilon_k$. Obviously, 
the width of the peak increases with increasing $\epsilon_k$. 
In Fig.\ref{fig1}e, we plot the full width at half maximum (FWHM) of the peak vs $\omega$ 
at  $\epsilon_k =0$ (EDC curve).
We see that {\it the width scales linearly with frequency over a wide frequency range}. 
In the inset to this figure we plot the FWHM of the peak in $k-$space 
at $\omega =0$ and varying temperatures (MDC curve). 
Again, the width clearly {\it scales linearly with $T$ over a substantial $T-$range}.
In Fig. 1f, we plot the velocity measured in the
 EDC dispersion 
 $v^*_F = v_F/(1 + d\Sigma^{\prime}_k (\omega)/d\omega)$ vs frequency. 
At $\omega =0$,
$v^*_F = v_F/(1 + \lambda)$. 
At larger frequencies, 
the  renormalization factor decreases due to the flattening of 
$\Sigma^{\prime}$, and $v^*_F$ gradually approaches $v_F$.  

\begin{figure}[tbp]
\centerline{\epsfxsize=3.5in \epsfysize=3.75in
\epsffile{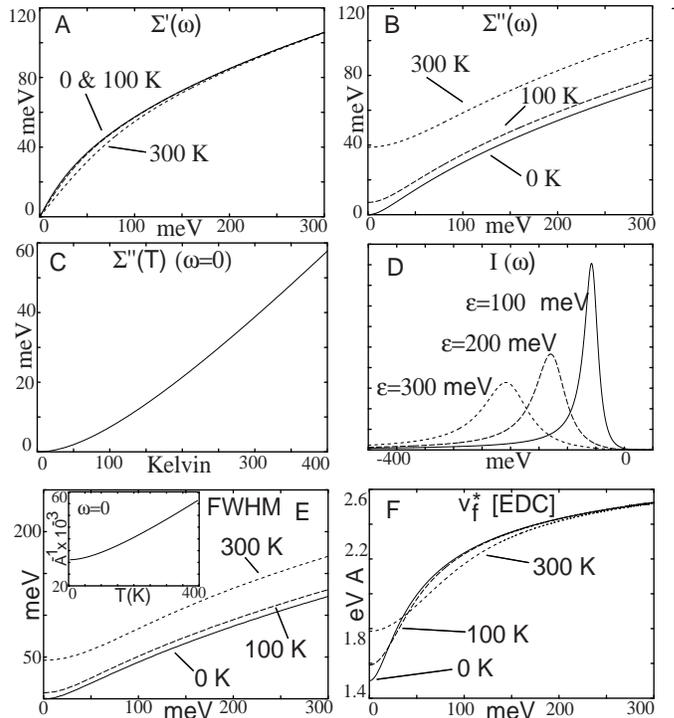}}
\caption{
Theoretical results for the fermionic self-energy, Eq.~\protect\ref{z}, 
and the photoemission intensity.
For definiteness, we used $\lambda =1$, $\omega_{sf} = 20meV$ and
$v_F = 3 eV A$ (this yields $\xi \approx 9 A$). 
Figs. a and b - $\Sigma^{\prime}(\omega)$ and $\Sigma^{\prime\prime}(\omega)$
respectively at various $T$, indicated on the figures. 
Fig. c - $\Sigma^{\prime\prime}(T)$ at
$\omega =0$. Fig. d - the photoemission intensity vs $\epsilon_k$.
Fig. e- the FWHM of the photoemission peak vs $\omega$ at $\epsilon_k =0$ 
(ECD curve). Inset -  the FWHM of the peak in $k-$space 
at $\omega =0$ vs $T$ (MDC curve).
Fig. f - the velocity of the EDC dispersion 
$v^*_F = v_F/(1 + d\Sigma^{\prime}/d\omega)$ vs $\omega$.}
\label{fig1}
\end{figure}

We now discuss how these results are modified 
away from the hot spots. 
For this we notice that at finite 
${\tilde k} = k - k_{hs}$,  Eq. (\ref{z}) can
be reduced to the same form 
as at a hot spot, if we introduce the $k$ dependent 
coupling $\lambda_k$ and  spin fluctuation 
frequency $\omega_{sf} (k)$ via
\begin{equation}
\lambda_k = \lambda/(1 + ({\tilde k} \xi)^2)^{1/2},~~ \omega_{sf} (k) = \omega_{sf} (1 +  ({\tilde k} \xi)^2)
\label{def}
\end{equation}   
We see therefore that away from the hot spots, 
the effective coupling gets smaller, 
and the crossover frequency $\omega_{sf} (k)$ increases. Still,
however, at frequencies/temperatures which exceed 
$\omega_{sf} (k)$, $\Sigma^{\prime \prime}_k (\omega, T)$ is linear in both frequency and temperature. Obviously then,  for intermediate 
$\omega$,  changing the momentum along
the Fermi surface only affects the overall slopes of 
$\Sigma^{\prime \prime}_k (\omega)$ vs $\omega$ and $T$. 
Furthermore, 
 at optimal doing $\xi \sim (1-2) a$,
 where $a \sim 3A$ is a $Cu-Cu$ distance, and 
$|{\tilde k}_0| \sim 0.3\pi/a$. 
Accordingly, $\lambda_k$ changes between $\lambda$ and 
$(0.7-0.4)\lambda$ between $k_{hs}$ 
and $k_{diag}$, i.e., the change in the coupling constant is not dramatic.

\begin{figure}[tbp]
\centerline{\epsfxsize=3.5in \epsfysize=3.5 in
\epsffile{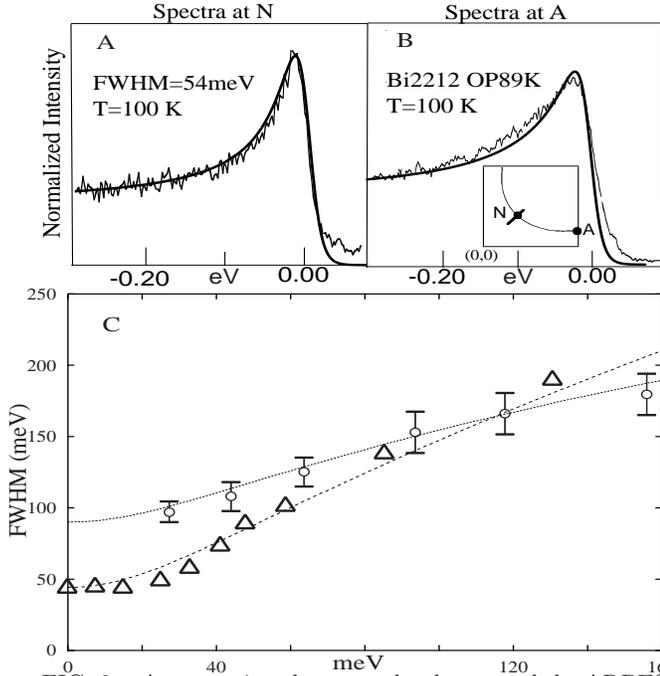}}
\caption{
A comparison between the theory and the ARPES data. a,b. A fit of Argonne 
data~\protect\cite{kam} for 
$k \sim k_{hs}$ (a) and $k = k_{diag}$ (b);
c. Fits of the FWHM of the EDC curves to the theory. The
circles with error bars are data from \protect\cite{camp} taken at
115 K and the triangles from \protect\cite{peter} taken at 90 K. 
The two theoretical curves are for different coupling constants (see text)}
\label{fig2}
\end{figure}
This analysis is indeed an approximate one, 
and care must be taken when applying it to 
momenta about ${\bf k}_F = {\bf k}_{diag}$ where $\phi_0 = \pi$,
and Eq. (\ref{z}) is, strictly speaking, unapplicable.  
This complication certainly affects the fermionic self-energy at the smallest
frequencies. However, we verified numerically that for 
frequencies $\omega \geq \omega_{sf} (k_{diag})$, the dominant contribution to
$\Sigma^{\prime \prime}_{k_{diag}} (\omega)$ comes from fermions 
away from diagonal,
 for which $\phi_0$ can still be approximated by a constant. 
We  therefoe will use Eq. (\ref{z}) for momenta both near the hot spots 
and along the zone diagonals with the understanding that in the latter case, 
the result has to be modified at the lowest frequencies.

In Fig. \ref{fig2}, we compare our theoretical results with the 
ARPES data for optimally doped $Bi2212$. In Figs.\ref{fig2} a and b,
we fit ARPES data taken by the Argonne group~\cite{kam} to our
formulas for 
$k \sim k_{diag}$ and $k = k_{hs}$.  The values 
$\lambda_{k_{diag}} = 1$, $\omega_{sf} (k_{diag}) = 20 meV$, 
and $\lambda_{k_{hs}} = 2$, $\omega_{sf} (k_{hs}) =5 meV$. 
respectively were used.
The value of $\lambda_{k_{hs}}$ agrees with our previous study of the position 
of the resonance peak at optimal doping~\cite{acs}. 
To account for the background, we added a momentum and frequency independent 
damping  $\gamma = 70 meV$ to  $\Sigma^{\prime \prime} (\omega)$. 
This damping likely comes from impurity scattering, but this is
not well understood and requires further study.  

In Fig. \ref{fig2}c, we compare the theoretical slope of the FWHM of the 
EDC curve with the Argonne data for the deviations from the Fermi surface 
along the zone diagonal.  Again we added a constant damping  $\gamma=70 meV$
to $\Sigma^{\prime \prime} (\omega)$. We see that the agreement is quite impressive. 
The second line in Fig. \ref{fig2}c is our fit to the Brookhaven~\cite{peter}
and Stanford~\cite{shen} data, which also yields a linear in frequency FWHM, but 
with a different slope. To fit this data, we just have to use 
$\lambda_{k_{diag}} \approx 2.5$. This value of the coupling 
is also quite reasonable. A slightly diffent $\gamma$ (55 meV)
was used for the second line.  We also  verified that the 
results for the FWHM are largely insensitive to changes in $\omega_{sf}$
which for Fig. \ref{fig2}c is 20 meV.

ARPES lineshapes at $k_{hs}$ and $k_{diag}$ have previously
been fitted using the marginal Fermi liquid phenomenology by
$\Sigma^{\prime \prime}_k (\omega) = A \omega + B_k$ with $k-$independent 
prefactor $A$~\cite{va}.
Our analysis shows that such data can equally 
well be fitted by   $\Sigma^{\prime \prime}_k (\omega) = A_k \omega + B$ with a 
moderately $k-$dependent $A_k$ and constant $B$. 
We emphasize, however, that the theoretical reasoning  
is qualitatively different in the two approaches.  
 
We now turn to the computation of the conductivity 
$\sigma (\omega) =  i \Pi (\omega)/\omega$, where $\Pi (\omega)$ is a current-current correlator.
Diagramatically, $\Pi (\omega)$ is given by a particle-hole bubble with 
$d \epsilon_k/dk$ in the vertices. In our calculations, the typical momenta are
comparable to $k_F$, and hence vertices can be treated as constants.
Also, since the self-energy, Eq. (\ref{z}),
depends predominantly on frequency,  vertex corrections to the 
particle-hole bubble (related to $d\Sigma/dk$ by the Ward identity) 
do not change the physics and can be safely neglected. 
For $k$ near the hot spots, the neglect of vertex corrections can also be justified by the
argument that
the velocities at $k$ and $k+Q$ are nearly orthogonal, and hence the
transport lifetime is the same as the conventional lifetime.  

Substituting the expressions for the self-energy into the particle-hole 
bubble and expanding, as before, to linear order in deviations from
the Fermi surface, one can explicitly perform the integration over $\epsilon_k$ and obtain
\begin{equation}
\Pi (\omega_m)\! =\! \frac{\omega^2_{pl}}{4}  T \!\sum_n\! \int \!d\tilde k 
\frac{\Theta (\omega_n + \omega_m) - \Theta (\omega_n)}{\omega_m + \Sigma_k (\omega_n + \omega_m) -\Sigma_k (\omega_n)} \nonumber
\end{equation}
where the momentum integration is along the Fermi surface, 
$\Sigma_k (\omega_n)$ is given by (\ref{z}),
 $\Theta(x)$ is a Theta-function, 
%$ =0 $ if $x<0$ and $\Theta(x) =1$ if $x>0$, 
and $\omega_{pl}$ is the plasma frequency. We used
$\omega_{pl}  \sim 1.2\times 10^4 cm^{-1}$, similar to that in~\cite{basov}.

The $k-$dependence of $\Sigma_k (\omega)$ 
emerges through the momentum dependence of 
$\lambda_k$ and $\omega_{sf} (k)$. As this dependence
is not dramatic  at optimal doping, 
it should not substantially affect the frequency 
dependence of $\Pi (\omega)$. To make the computations more transparent, we
 replace  
$\lambda_k$ and $\omega_{sf} (k)$ by averaged, $k$-independent  
values. 

\begin{figure}[tbp]
\centerline{\epsfxsize=3.5in \epsfysize=2.5in
\epsffile{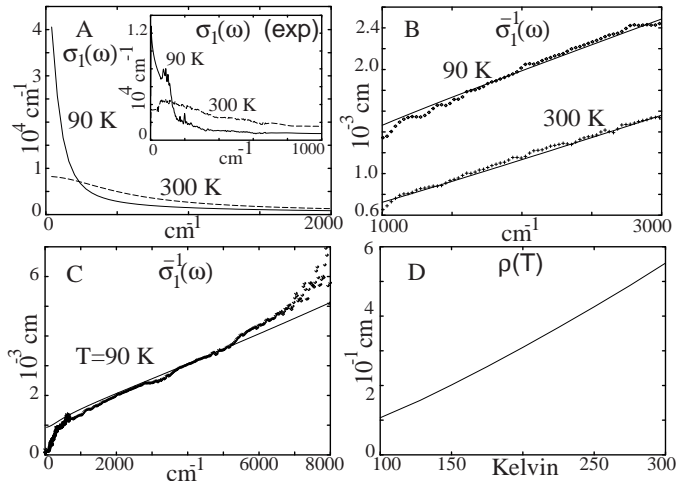}}
\caption{
Optical conductivity at optimal doping. 
For definiteness we used 
$\lambda =2$, $\omega_{sf} = 5 meV$
a. The theoretical results for $\sigma_1 (\omega)$.  The experimental 
data is from~\protect\cite{basov} 
is shown in the inset; b. $\sigma^{-1}_1$ vs $\omega$ for $T=90K$ and 
$T=300K$. The lines are theoretical curves,
 and the points are the experimental data; c. $\sigma^{-1}_1 (\omega)$ in
  the extended frequency range. The experimental data is 
from Ref \protect\cite{basov}; 
 e)the resistivity $\rho(T)$ vs $T$.}
\label{fig3}
\end{figure}

We computed $\Pi^{\prime \prime} (\omega)$ 
by using a spectral representation, and 
used a Kramers-Kronig transformation to obtain $\Pi^{\prime} (\omega)$. 
In Fig. \ref{fig3}a   
we present  our theoretical results for 
$\sigma_1 (\omega) = Re \sigma (\omega)$ at T=90 and 300 K. 
The inset of this figure contains experimental data 
from Ref~\cite{basov}.  In Fig. \ref{fig3}b,
we plot $1/\sigma_1 (\omega)$ for frequencies above $1000 cm^{-1}$
and compare it with the data.  We see that the theoretical 
$1/\sigma_1 (\omega)$ {\it is linear in frequency}.
 This fully agrees with the data. 
Furthermore, we found that  
$\lambda =2$, compatible to that which we used for 
the spectral function, yields  a perfect  fit
to the measured slopes of $1/\sigma_1 (\omega)$ both at 
$90K$ and $300K$. We consider this agreement a 
 strong indication that spin-fluctuation mechanism 
captures the essential physics of the cuprates. 
To match the values of the conductivity, we, however, again 
have to add a frequency independent 
(but temperature dependent) constant to $1/\sigma_1$.
 As roughly, 
$1/ \sigma_1 (\omega) \propto  \Sigma^{\prime \prime} (\omega)$, this 
 procedure is qualitatively similar (although not quite equivalent) 
 to adding a constant to $\Sigma^{\prime \prime} (\omega)$.
Similar procedure has been used in Ref.~\cite{im}.
In Fig. \ref{fig3}c we plot  our $1/\sigma_1 (\omega)$ in the extended 
frequency range, up to $8000 cm^{-1}$. We see 
that the theoretical 
linear behavior of $1/\sigma_1$ extends up to very large frequencies 
of $7000 cm^{-1}$, where $\Sigma^{\prime \prime} (\omega)$  already curves down
from a linear behavior.  
This extension of the linear regime to very high frequencies is also  
consistent with the experimental data of~\cite{basov}. 
%In Fig. \ref{fig3}d, we present the results for 
%$1/\tau (\omega) = Re [1/\sigma (\omega)]$ 
%together with the experimental data from ~\cite{basov}.
%We again see that the agreement is quite good.
Finally, in  Fig. \ref{fig3}e, 
we present the result for the resistivity $\rho (T)$.
We see that $\rho (T)$ is linear in $T$ above $100 K$. 
This  is indeed a consequence of the linearity
of $\Sigma^{\prime \prime} (T, \omega =0)$.

To summarize, in this paper we used the spin-fluctuation approach to
calculate the photoemission intensity, optical conductivity
and resistivity in the normal state of the cuprates near optimal doping. 
We found that the fermionic self-energy is linear in both $\omega$ and $T$
in a wide range of frequencies and temperatures. This gives rise to a
linear frequency dependence of the inverse conductivity %and $1/\tau$, 
and to a linear temperature dependence of the resistivity. 
We performed {\it quantitative} comparisons with the experimental data and found  
near perfect matches of the slopes for both 
the ARPES linewidth and the conductivity. 
We view the results as a strong indication that 
the interaction between low-energy fermions and their 
spin collective degrees of freedom is the 
dominant scattering mechanism in the cuprates. The open issue 
%which wedid not discuss in this paper 
is whether the spin-fluctuation scenario 
is capable of explaining the differences between the temperature dependence
of the diagonal and Hall conductivities~\cite{im,david,hall}. 
This study is currently under way. 

It is our pleasure to thank D. Basov, G. Blumberg, J.C. Campuzano
P. Johnson, M. Norman, J. Schmalian, 
O. Tschernishev and V. Yakovenko for useful
discussions. We are also 
thankful to D. Basov for sending us the unpublished experimental data.
 The research was supported by NSF DMR-9979749 (Ar. A and A. Ch.)
and by NSF DMR-96-32527 (R.H.).

\end{document}